\documentstyle[twocolumn,prb,aps]{revtex}

\begin{document}     
\twocolumn[
\hsize\textwidth\columnwidth\hsize\csname@twocolumnfalse\endcsname

\title{Charge dynamics in the Mott insulating phase of the ionic Hubbard model}
\author{A.A. Aligia}
\address{Centro At\'{o}mico Bariloche and Instituto Balseiro, \\
Comisi\'on Nacional de Energ\'{\i}a At\'{o}mica, \\
8400 Bariloche, Argentina}
\maketitle

\begin{abstract}
I extend to charge and bond operators the transformation that maps the
ionic Hubbard model at half filling onto an effective spin Hamiltonian.
Using the transformed operators I calculate the amplitude of the charge 
density wave in different dimensions D. In 1D, the charge-charge correlations at
large distance $d$ decay as $d^{-3}\ln ^{-3/2}d$, in spite of the finite
charge gap, due to remaining charge-spin coupling.
Bond-bond correlations decay as $(-1)^{d}d^{-1}\ln ^{-3/2}d$ as in the usual
Hubbard model.
\end{abstract}

\draft


]


\section{Introduction}

\label{Introduction} The ionic Hubbard model (IHM) has been proposed in the
80's for the description of the neutral-ionic transition in mixed-stack
charge-transfer organic crystals. \cite{hub,nag} In the 90's the interest on
the model increased due to its potential application to ferroelectric
perovskites. \cite{ega,res,ort,res2,gid,fab,tor} The model in any bipartite
lattice can be written as: 
\begin{eqnarray}
H &=&H_{0}+H_{t};\text{ }H_{0}=\frac{\Delta }{2}\sum_{i\sigma
}(-1)^{i}n_{i\sigma }+U\sum_{i}n_{i\uparrow }n_{i\downarrow }  \nonumber \\
&&H_{t}=-t\sum_{\delta \sigma }c_{i+\delta \sigma }^{\dagger }c_{i\sigma },
\label{h}
\end{eqnarray}
where $i+\delta $ denote the nearest neighbors of site $i$, and odd $i$
correspond to the sublattice with on-site energy $-\Delta /2$.

At zero temperature, in the strong coupling limit $t=0$, the system is in
the Mott insulating (MI) phase for $U>\Delta $ (all sites are singly
occupied). and in the bond insulating (BI) phase for $U<\Delta $ (all sites
with odd $i$ doubly occupied). In one dimension (1D), for finite $t$, field
theory calculations pointed out the existence of a third intermediate phase,
the spontaneous dimerized insulator (SDI) between the other two. \cite{fab}
For fixed $\Delta $, as $U$ increases, first a transition at $U_{c}$
(involving mainly charge degrees of freedom) between the BI and the SDI
takes place, and then for $U_{s}>$ $U_{c}$, the spin gap closes as in usual
Kosterlitz-Thouless transitions. The physics has some similarities to that
of the extended Hubbard model for small $U$ and nearest-neighbor repulsion $%
V $. \cite{tor,nak,tsu}

The phase diagram of the 1D IHM has been calculated numerically using the
method of jumps of Berry phases, \cite{tor} which in this case coincides
with the method of crossings of appropriately chosen excited levels. \cite
{tor,nak} The charge Berry phase is a measure of the polarization and
therefore detects the ferroelectric charge transition, \cite{res,ort,res2}
while the spin Berry phase jumps at the point where the spin gap closes. 
\cite{epl} For small $t$, $U_{s}-$ $U_{c}\cong 0.6t$ was obtained for the
width of the SDI phase. \cite{tor}

Due to the fact that the spin gap is exponentially small for $U$ smaller but
near $U_{s}$, direct numerical calculation of it or of appropriate
correlation functions are unable to find a sharp transition at $U_{s}$. \cite
{wil,brune} The same difficulty happens in the Hubbard chain with correlated
hopping, \cite{topo} where the existence and position of the transition is
confirmed by field theory \cite{jap,bos} and exact \cite{afq} results. In
the IHM some controversy remains about the existence of $U_{s}$ and the
nature of the phase for $U>U_{c}$ in the case of absence of a second
transition. Wilkens and Martin \cite{wil} suggested that the MI phase does
not exist (the spin gap remains open for any finite $U$). This is in
contradiction with the strong coupling expansion for $t\ll U-\Delta $, which
maps the IHM onto an effective spin Hamiltonian $\tilde{H}$ with closed spin
gap. \cite{nag,brune} Wilkens and Martin \cite{wil} argued that since $%
\tilde{H}$ has an additional symmetry with respect to $H$ (translation to a
nearest neighbor $T_{\delta }$), $\tilde{H}$ seems to have lost part of the
physics of $H$, which might be essential.

Recently, a detailed study of different observables of the 1D IHM was made,
using density-matrix renormalizarion group (DMRG), including a careful
finite-size scaling analysis. \cite{man} The results indicate the present of
two transitions and are consistent with the phase diagram found using
topological transitions, \cite{tor} although a smaller width of the SDI
phase is suggested. Surprisingly, the authors find a power law decay of
charge-charge correlation functions in the MI phase, in spite of a 
the presence of a charge gap $\sim U-\Delta$.

Here I report calculations of charge expectations values, and in one
dimension, the long distance behavior of charge-charge and bond-bond
correlation functions using the mapping to a spin Hamiltonian, valid for $%
t\ll U-\Delta $. This might seem surprising at first sight, since charge
fluctuations are frozen in $\tilde{H}$. However, the relevant information is
contained in the transformed operators. This has some similarities with the
case of the cuprates, in which Cu and O contributions to the photoemission
spectra were successfully calculated using effective one-band models. \cite
{sf,fei,opt,ero} Also, while a $t-J$ model is enough to explain the observed
dispersion in photoemission measurements in Sr$_{2}$CuO$_{2}$Cl$_{2}$, the
mapping of the operators \cite{esk,lema} is crucial to explain the observed
intensities. \cite{lema}

The transformation on the operators is performed in Section II. Section III\
contains the results for expectation values of different quantities, and
Section IV is a discussion.

\section{The canonical transformation}

In this Section, I use the canonical transformation plus projection onto the
low-energy subspace that maps $H$ into a spin Hamiltonian $\tilde{H}$, to
transform charge and bond operators in one direction ($\delta =\pm 1$)

\begin{eqnarray}
n_{i} &=&n_{i\uparrow }+n_{i\downarrow },  \nonumber \\
b_{i} &=&\sum_{\delta \sigma }\delta (c_{i+\delta \sigma }^{\dagger
}c_{i\sigma }+{\rm H.c.}),  \label{op}
\end{eqnarray}
and discuss some symmetry properties. For our purposes, it is enough to work
up to second order in $t/(U-\Delta )$. Thus

\begin{equation}
\tilde{H}=Pe^{-S}He^{S}P=P(H+[H,S]+\frac{1}{2}[[H,S],S]+...)P,  \label{can}
\end{equation}
where $P$ is the projector over the ground state subspace of $H_{0}$, in
which $n_{i}\equiv 1$ for all $i$, and $S$ is chosen in such a way that
linear terms in the hopping term $H_{t}$ are eliminated: $H_{t}+[H_{0},S]=0$%
. Taking matrix elements of this equation between eigenstates of $H_{0}$,
one obtains:

\begin{equation}
\langle n|S|m\rangle =\frac{\langle n|H_{t}|m\rangle }{E_{m}-E_{n}},
\label{s}
\end{equation}
where $E_{j}$ is the energy of the eigenstate $|j\rangle $ of $H_{0}$. From
Eqs. (\ref{can}) and (\ref{s}), proceeding in a similar way as done below,
one obtains the known result: \cite{nag}

\begin{equation}
\tilde{H}=\frac{4t^{2}U}{(U^{2}-\Delta ^{2})}\sum_{\langle ij\rangle }({\bf S%
}_{i}\cdot {\bf S}_{j}-\frac{1}{4}).  \label{hh}
\end{equation}
The most important correction to $\tilde{H}$ in higher order is a
next-nearest-neighbor exchange of order $t^{4}$, which does not affect the
physics for $t\ll U-\Delta $. \cite{nag,brune}

The effective Hamiltonian $\tilde{H}$ to all orders in $t$ is invariant
under a nearest-neighbor translation $T_{\delta }$, while $H$ is not.\cite
{nag} This is simply a consequence of the fact that to all orders $\tilde{H}$
is a purely spin Hamiltonian, since the charge degrees of freedom are frozen
($n_{i}\equiv 1$ for all $i$). Then $\tilde{H}$ is invariant under the
electron-hole transformation $T_{eh}:c_{i\sigma }^{\dagger }\rightarrow
\sigma c_{i\bar{\sigma}}$, which leaves invariant all spin operators. Since
the original Hamiltonian $H$ is invariant under the product $T_{\delta
}T_{eh}$, then $\tilde{H}$ should also be invariant under $T_{\delta }$.
However other transformed operators, like $\tilde{n}_{i}$ below, are not
invariant under $T_{\delta }.$

The transformed charge operator is:

\begin{eqnarray}
\tilde{n}_{i} &=&Pe^{-S}n_{i}e^{S}P\cong P(n_{i}+[n_{i},S]+\frac{1}{2}%
[[n_{i},S],S])P  \nonumber \\
&=&1+P(SS-Sn_{i}S)P.  \label{naux}
\end{eqnarray}
In the second equality I used the fact that $S$ applied to any state of the
ground state manifold of $H_{0}$, gives an excited state. The only terms of $%
-PSn_{i}SP$ which do not cancel with $PSSP$ are those in which $n_{i}=0$ or $%
n_{i}=2$ after the first application of $S$. Using Eq. (\ref{s}):

\begin{eqnarray}
\tilde{n}_{i} &=&1+\sum_{\delta \sigma \sigma ^{\prime }}P[-\left( \frac{t}{%
U-(-1)^{i}\Delta }\right) ^{2}c_{i\sigma ^{\prime }}^{\dagger }c_{i+\delta
\sigma ^{\prime }}c_{i+\delta \sigma }^{\dagger }c_{i\sigma }  \nonumber \\
&&+\left( \frac{t}{U+(-1)^{i}\Delta }\right) ^{2}c_{i+\delta \sigma
}^{\dagger }c_{i\sigma }c_{i\sigma ^{\prime }}^{\dagger }c_{i+\delta \sigma
^{\prime }}]P.  \label{naux2}
\end{eqnarray}
The first (negative) term correspond to charge $n_{i}=0$ in the intermediate
state and the excitation energy is $E_{n}-E_{m}=U+(-1)^{i+\delta }\Delta
/2-(-1)^{i}\Delta /2=U-(-1)^{i}\Delta $. The sums over spins can be
transformed as follows:

\begin{eqnarray}
\sum_{\sigma \sigma ^{\prime }}Pc_{j\sigma }^{\dagger }c_{i\sigma
}c_{i\sigma ^{\prime }}^{\dagger }c_{j\sigma ^{\prime }}P &=&\sum_{\sigma
\sigma ^{\prime }}Pc_{j\sigma }^{\dagger }c_{j\sigma ^{\prime }}(\delta
_{\sigma \sigma ^{\prime }}-c_{i\sigma ^{\prime }}^{\dagger }c_{i\sigma })P 
\nonumber \\
&=&\sum_{\sigma }Pn_{j\sigma }(1-n_{i\sigma
})P-S_{j}^{+}S_{i}^{-}-S_{j}^{-}S_{i}^{+}=  \nonumber \\
&=&\frac{1}{2}-2{\bf S}_{i}\cdot {\bf S}_{j}.  \label{ss}
\end{eqnarray}
And replacing this in Eq. (\ref{naux2}):

\begin{equation}
\tilde{n}_{i}=1-(-1)^{i}\frac{2U\Delta t^{2}}{(U^{2}-\Delta ^{2})^{2}}%
\sum_{\delta }(1-4{\bf S}_{i}\cdot {\bf S}_{i+\delta }).  \label{nt}
\end{equation}
This equation relates a charge operator with a purely spin operator.
Although this might seem surprising at first sight, it is clear that in the
strong coupling limit, virtual charge fluctuations are inhibited only for a
ferromagnetic alignment of the spins. This shows that for finite $U$, some
charge degrees of freedom remain coupled with spin, in contrast to the usual
Hubbard model ($\Delta =0).$

Using Eq. (\ref{op}), to leading order in $t$, the transformed bond operator
is:

\begin{eqnarray*}
\tilde{b}_{i} &=&Pe^{-S}b_{i}e^{S}P\cong P[b_{i},S]P \\
&=&\sum_{\delta \sigma \sigma ^{\prime }}\delta P[\frac{t}{U+(-1)^{i}\Delta }%
c_{i+\delta \sigma }^{\dagger }c_{i\sigma }c_{i\sigma ^{\prime }}^{\dagger
}c_{i+\delta \sigma ^{\prime }} \\
&&+\frac{t}{U-(-1)^{i}\Delta }c_{i\sigma ^{\prime }}^{\dagger }c_{i+\delta
\sigma ^{\prime }}c_{i+\delta \sigma }^{\dagger }c_{i\sigma }+{\rm H.c.}]P.
\end{eqnarray*}
The first term correspond to acting first with $S$ creating an excitation
with a doubly occupied site at $i$, and an empty site at $i+\delta $, and
then with $b_{i}$, restoring occupation 1 at each site. Using Eq. (\ref{ss}%
), I obtain:

\begin{equation}
\tilde{b}_{i}=\frac{-8tU}{U^{2}-\Delta ^{2}}\sum_{\delta }\delta {\bf S}%
_{i}\cdot {\bf S}_{i+\delta }.  \label{bt}
\end{equation}
The second member is a local measure of the asymmetry between the spin
``bonds'' involving site $i$ in the direction of $\delta $. Here there is no
essential difference with the result for the ordinary Hubbard model.

\section{Observables and correlation functions}

For $t\ll U-\Delta $, using Eq. (\ref{nt}) the amplitude of the charge
density wave is given by:

\begin{eqnarray}
A &=&|\langle n_{i}-n_{i+\delta }\rangle _{H}|=|\langle \tilde{n}_{i}-\tilde{%
n}_{i+\delta }\rangle _{\tilde{H}}|  \nonumber \\
&=&2a\sum_{\delta }(1-4\langle {\bf S}_{i}\cdot {\bf S}_{i+\delta }\rangle _{%
\tilde{H}}),  \label{aa}
\end{eqnarray}
where the subscript in the expectation values indicates the Hamiltonian with
which they are calculated and

\begin{equation}
a=\frac{2U\Delta t^{2}}{(U^{2}-\Delta ^{2})^{2}}.  \label{a}
\end{equation}

In 1D, from the Bethe ansatz solution $1/4-\langle {\bf S}_{i}\cdot {\bf S}%
_{i+\delta }\rangle _{\tilde{H}}=\ln 2\cong 0.69$. \cite{bah}. This gives $%
A\cong 11.09a$. This result has also been obtained using Hellman-Feynman
theorem and is in very good agreement with DMRG results. \cite{man} For the
square lattice $\langle {\bf S}_{i}\cdot {\bf S}_{i+\delta }\rangle _{\tilde{%
H}}\cong -0.3347$, \cite{cal} and then $A\cong 18.7a$. In three dimensions,
spin waves is a good aproximation and gives $\langle {\bf S}_{i}\cdot {\bf S}%
_{i+\delta }\rangle _{\tilde{H}}\cong -0.30$, \cite{kitt} what leads to $%
A\cong 26.3a.$

The charge-charge correlation function is given by:

\begin{equation}
C_{d}=\langle n_{i}n_{i+d}\rangle _{H}-\langle n_{i}\rangle _{H}\langle
n_{i+d}\rangle _{H}.  \label{ccc}
\end{equation}
To leading order, I can approximate $\langle n_{i}n_{i+d}\rangle _{H}\cong
\langle \tilde{n}_{i}\tilde{n}_{i+d}\rangle _{\tilde{H}}$. The neglected
term (see Eq. (\ref{naux}) ) $\langle
Pe^{-S}n_{i}e^{S}(1-P)e^{-S}n_{i+d}e^{S}P\rangle _{\tilde{H}}$ involves
charge fluctuations across the energy gap and therefore should lead to an
exponentially decaying contribution, as in the usual Hubbard model. Using
Eq. (\ref{nt}) in the form $\tilde{n}_{i}=1+(-1)^{i}4a{\bf S}_{i}\cdot {\bf S%
}_{i+\delta }-z(-1)^{i}a+O(t^{4})$, where $z$ is the number of nearest
neighbors, and dropping the subscript $\tilde{H}$, one obtains to leading
order in $t$:

\begin{eqnarray}
C_{d} &=&16a^{2}(-1)^{d}\sum_{\delta \delta ^{\prime }}[\langle ({\bf S}%
_{i}\cdot {\bf S}_{i+\delta })({\bf S}_{i+d}\cdot {\bf S}_{i+d+\delta
})\rangle  \nonumber \\
&&-\langle {\bf S}_{i}\cdot {\bf S}_{i+\delta }\rangle \langle {\bf S}%
_{i+d}\cdot {\bf S}_{i+d+\delta }\rangle ].  \label{cca}
\end{eqnarray}
Note that the terms $O(t^{4})$ in $\tilde{n}_{i}$ cancel exactly in Eq.(\ref
{ccc}). Thus, it is enogh to include terms up to order $t^{2}$ in $\tilde{n}%
_{i}$ to obtain the result up to order $t^{4}$ in $C_{d}$. Using symmetry, I
can write: 
\begin{eqnarray}
\langle ({\bf S}_{i}\cdot {\bf S}_{i+\delta })({\bf S}_{i+d}\cdot {\bf S}%
_{i+d+\delta })\rangle &=&3\langle S_{0}^{z}S_{\delta
}^{z}S_{d}^{z}S_{d+\delta }^{z}\rangle  \nonumber \\
&&+6\langle S_{0}^{x}S_{\delta }^{x}S_{d}^{y}S_{d+\delta }^{y}\rangle ,
\label{cca2}
\end{eqnarray}
and similarly for the other term inside square brackets of Eq. (\ref{cca}).

In 1D, the leading power-law decay can be determined using bosonization
expressions for the XXZ model and correlation functions of the gaussian
model. \cite{gog}. However, at the isotropic point, there are important
logarithmic corrections. They can be calculated using expressions derived by
Giamarchi and Schulz.\cite{gia} Using $S_{l}^{z}\approx \cos (\pi l+\sqrt{2}%
\varphi (x))$ for the slowest decaying part of $S_{l}^{z}$, the operator
product expansion $S_{l}^{z}S_{l+1}^{z}\approx \cos (2\sqrt{2}\varphi (x))$,
and the results of section III B of Ref. \cite{gia}, I obtain:

\begin{equation}
\langle S_{0}^{z}S_{\delta }^{z}S_{d}^{z}S_{d+\delta }^{z}\rangle \approx
d^{-4}\ln ^{-2}d.  \label{zz}
\end{equation}
Using symmetry again:

\begin{eqnarray}
\langle S_{0}^{x}S_{\delta }^{x}S_{d}^{y}S_{d+\delta }^{y}\rangle &=&\frac{1%
}{8}\langle (S_{0}^{+}S_{\delta }^{-}+S_{0}^{-}S_{\delta
}^{+})(S_{d}^{+}S_{d+\delta }^{-}+S_{d}^{-}S_{d+\delta }^{+})\rangle 
\nonumber \\
&&-\langle S_{0}^{z}S_{\delta }^{z}S_{d}^{z}S_{d+\delta }^{z}\rangle .
\label{xy}
\end{eqnarray}
The first term in this equation turns out to be the dominant one, and
therefore I explain it in more detail. Performing a Jordan-Wigner
transformation from spin operators to fermions with annihilation operators $%
a_j$, going to the continuum limit using $a_j=i^j L(x)+(-i)^j R(x)$, with $%
x=ja$, and then bosonizing one gets:

\begin{eqnarray}
\sum_{\delta} (S_{j}^{+}S_{j+\delta }^{-}+{\rm H.c.}) \rightarrow
\sum_{\delta} ((a_{j}^{\dagger }a_{j+\delta}+{\rm H.c.})  \nonumber \\
\rightarrow \sum_{\delta}[i^\delta L^{\dagger }(x)L(x+a \delta) +
(-i)^\delta R^{\dagger }(x)R(x+a \delta) +  \nonumber \\
(-1)^j (-i)^\delta L^{\dagger }(x)R(x+a \delta) + (-1)^j i^\delta R^{\dagger
}(x)L(x+a \delta) +{\rm H.c.}]  \nonumber \\
\rightarrow 2i[L^{\dagger } \frac{\partial L}{\partial x} -R^{\dagger } 
\frac{\partial R}{\partial x} -(-1)^j L^{\dagger } \frac{\partial R}{%
\partial x} +(-1)^j R^{\dagger } \frac{\partial L}{\partial x}]+{\rm H.c.} 
\nonumber \\
= 2i[2L^{\dagger } \frac{\partial L}{\partial x} -2R^{\dagger } \frac{%
\partial R}{\partial x} +(-1)^j \frac{\partial}{\partial x}( R^{\dagger }L -
L^{\dagger }R)]  \nonumber \\
\rightarrow 4[(\frac{\partial \varphi}{\partial x})^2 +(\frac{\partial \theta%
}{\partial x})^2 ] +c (-1)^j \frac{\partial}{\partial x} \cos (\sqrt{2}%
\varphi)  \label{bos}
\end{eqnarray}
where $c$ is a nonuniversal constant of the order of one.

I would like to remark that the bosonic fields $\varphi $, $\theta $ are not
pure spin operators of the original Hamiltonian Eq. (\ref{h}), because they
come from operators already dressed by the transformation. The expression of
the transformed original fermions for $\Delta =0$ and any filling to lowest
order in $t/U$ can be found in Refs. \cite{esk,lema}.

The first two terms of the last member of Eq. (\ref{bos}) lead also to
contributions of order $1/d^4$ (without including logarithmic corrections).
Using \cite{gia}:

\begin{equation}
\langle :\cos (\sqrt{2}\varphi (x))::\cos (\sqrt{2}\varphi (y)):\rangle \sim 
\frac{1}{|x-y|\ln ^{3/2}|x-y|},  \label{cos}
\end{equation}
together with Eqs. (\ref{cca}) to (\ref{cos}), I finally obtain (except for
some factor of the order of one):

\begin{equation}
C_{d}\approx 48\frac{U^{2}\Delta ^{2}t^{4}}{(U^{2}-\Delta ^{2})^{4}}
d^{-3}\ln ^{-3/2}d.  \label{cc1d}
\end{equation}
Note that factors $(-1)^d$ cancel, and then in spite of the even-odd
oscillations in $\langle n_{j}\rangle$, these oscillations are absent in the
charge-charge correlation functions.

Concerning the bond-bond correlation functions, using Eqs. (\ref{bt}), (\ref
{cos}) and $\sum_{\delta }\delta {\bf S}_{i}\cdot {\bf S}_{i+\delta }\sim
\cos (\sqrt{2}\varphi (x))$, one obtains:

\begin{equation}
\langle b_{i}b_{i+d}\rangle \approx 64\frac{U^{2}t^{2}}{(U^{2}-\Delta
^{2})^{2}}(-1)^{d}d^{-1}\ln ^{-3/2}d.  \label{bb}
\end{equation}

These are much smaller than the spin-spin correlation functions, which
behave like $\langle S_{i}^{z}S_{i+d}^{z}\rangle \approx d^{-1}\ln ^{1/2}d$
at large distances, and have a larger prefactor.

\section{Discussion}

I have calculated several quantities of the ionic Hubbard model, using a
mapping to a spin Hamiltonian valid for $U>\Delta $ and $t\ll U-\Delta $. In
this limit the system is expected to be in the MI phase. This is predicted
to be the case in 1D, \cite{fab,tor} and in fact the absence of a spin gap
in the effective transformed Hamiltonian $\tilde{H}$ is consistent with this.%
\cite{nag,brune} Also, the bond-bond correlation functions decay faster than
the spin-spin ones at large distances, in contrast to what is expected in
the SDI.

By construction, the symmetry of the observables is independent of the basis
in which they are calculated, and this becomes clear after transformation of
the operators. For example, the charge operator $n_{i}$ is invariant under
translation of one lattice parameter $T_{\delta }$, while $H$ is not. These
symmetry properties are interchanged in the transformed operators $\tilde{n}%
_{i}$ and $\tilde{H}$, and the expectation value $\langle n_{i}\rangle
_{H}=\langle \tilde{n}_{i}\rangle _{\tilde{H}}$ is not invariant under $%
T_{\delta }$. As a consequence, for any finite $U$ there is a charge density
wave. Its amplitude for $t\ll U-\Delta $ is proportional to $t^{2}\Delta
U/(U^{2}-\Delta ^{2})^{2}$ but with a coefficient larger than 20 which
depends on dimension, and is near 53 in three dimensions.

The charge-charge correlation functions in 1D decay as $d^{-3}\ln ^{-3/2}d$
as a function of distance $d$ for large $d$ with a prefactor proportional to 
$t^{4}\Delta ^{2}$, in excelent agreement with recent DMRG calculations. 
\cite{man} This result for a finite charge gap, is in marked contrast to
the general behavior found in 1D systems and is a consequence of the fact
that coupling between charge and spin dynamics remains at low energies {\em %
in the original basis} (effective dressed charge and spin fields are
separated at low energies). This fact is at variance with the low-energy
spin-charge separation that takes place in translationally invariant models
with quite general nearest-neighbor interactions \cite{bos} (including the
extended Hubbard model with nearest-neighbor repulsion $V$ \cite{nak,tsu}
and the Hubbard model with correlated hopping \cite{nak,jap,bos}). An
important difference is that in weak coupling 
the charge-spin interaction is a relevant operator in the IHM. \cite{brune}

\section*{Acknowledgments}

I am indebted to R. Noack, A. Dobry and G. Japaridze for useful discussions.
This work was sponsored by PICT 03-06343 of ANPCyT. I am partially supported
by CONICET.


\begin{references}
\bibitem{hub}  J. Hubbard and J.B. Torrance, Phys. Rev. Lett. {\bf 47}, 1750
(1981).

\bibitem{nag}  N. Nagaosa and J. Takimoto, J. Phys. Soc. Jpn. {\bf 55}, 2735
(1986).

\bibitem{ega}  T. Egami, S. Ishihara, and M.Tachiki, Science {\bf 261}, 1307
(1993); Phys.Rev. B {\bf 49}, 8944 (1994).

\bibitem{res}  R. Resta and S. Sorella, Phys. Rev. Lett. {\bf 74}, 4738
(1995).

\bibitem{ort}  G. Ortiz, P. Ordej\'{o}n, R.M. Martin, and G. Chiappe, Phys.
Rev. B {\bf 54}, 13 515 (1996); references therein.

\bibitem{res2}  R. Resta and S. Sorella, Phys. Rev. Lett. {\bf 82}, 370
(1999).

\bibitem{gid}  N. Gidopoulos, S. Sorella, and E. Tosatti, Eur. Phys. J. B 
{\bf 14}, 217 (2000).

\bibitem{fab}  M. Fabrizio, A.O. Gogolin, and A.A. Nersesyan, Phys. Rev.
Lett {\bf 83}, 2014 (1999).

\bibitem{tor}  M.E. Torio, A.A. Aligia, and H.A. Ceccatto, Phys. Rev. B {\bf %
64}, 121105 (R) (2001).

\bibitem{nak}  M. Nakamura, J. Phys. Soc. Jpn. {\bf 68}, 3123 (1999); Phys.
Rev. B {\bf 61}, 16 377 (2000).

\bibitem{tsu}  M. Tsuchiizu and A. Furusaki, Phys. Rev. Lett {\bf 88},
056402 (2002).

\bibitem{epl}  A.A. Aligia, Europhys. Lett. {\bf 45}, 411 (1999).

\bibitem{wil}  T. Wilkens and R.M. Martin, Phys. Rev. B {\bf 63}, 235108
(2001).

\bibitem{brune}  A.P. Kampf, M. Sekania, G.I. Japaridze, and P. Brune, J.
Phys. C (in press).

\bibitem{topo}  A.A. Aligia, K. Hallberg, C.D. Batista, and G. Ortiz, Phys.
Rev B {\bf 61}, 7883 (2000).

\bibitem{jap}  G.I. Japaridze and A.P. Kampf, Phys. Rev B {\bf 59}, 12 822
(1999).

\bibitem{bos}  A.A. Aligia and L. Arrachea, Phys. Rev. B {\bf 60}, 15 332
(1999).

\bibitem{afq}  L. Arrachea, A.A. Aligia and E. Gagliano, Phys. Rev. Lett. 
{\bf 76}, 4396 (1996); references therein.

\bibitem{man}  S.R. Manmana, V. Meden, R.M. Noack, and K. Sch\"{o}nhammer,
cond-mat/0307741

\bibitem{sf}  C.D. Batista and A.A. Aligia, Phys. Rev. B {\bf 47}, 8929
(1993).

\bibitem{fei}  L.F. Feiner, Phys. Rev. B 48, 16 857 (1993).

\bibitem{opt}  M.E. Simon, A.A. Aligia, and E.R. Gagliano, Phys. Rev. B {\bf %
56}, 5637 (1997).

\bibitem{ero}  J. Eroles, C.D. Batista, and A.A. Aligia, Phys. Rev. B {\bf 59%
}, 14092 (1999).

\bibitem{esk}  H. Eskes, A.M. Oles, M.B.J. Meinders and W. Stephan, Phys.
Rev. B {\bf 50} 17 980 (1994); references therein.

\bibitem{lema}  F. Lema and A.A. Aligia, Phys. Rev. B {\bf 55}, 14092
(1997); Physica C {\bf 307}, 307 (1998).

\bibitem{bah}  See for example J. des Cloizeaux and J.J. Pearson, Phys. Rev. 
{\bf 128}, 2131 (1962).

\bibitem{cal}  M. Calandra and S. Sorella, Phys. Rev. B {\bf 57}, 11446
(1998).

\bibitem{kitt}  C. Kittel, {\it Quantum Theory of Solids}, (John Wiley \&
Sons, New York, 1987).

\bibitem{gog}  A.O. Gogolin, A.A. Nersesyan, and A.M. Tsvelick, {\it %
Bosonization and strongly correlated systems} (University Press, Cambridge,
1998).

\bibitem{gia}  T. Giamarchi and H.J. Schulz, PRB {\bf 39}, 4620 (1989).
\end{references}
\end{document}